\documentclass[twocolumn,letterpaper,aps,prc,superscriptaddress,nofootinbib,showpacs,floatfix]{revtex4-1}
\usepackage{graphicx}
\usepackage{comment}
\usepackage{setspace}
\usepackage{amsmath}
\usepackage{amssymb}
\usepackage[textwidth=16cm,textheight=22cm]{geometry}
\usepackage{color}
\usepackage{indentfirst}
\usepackage{xspace}
\usepackage{hyperref}
\usepackage{verbatim}
\usepackage{epstopdf}

\usepackage{lineno}
{\large }

\begin{document}
\title{Separating Diffractive and Non-Diffractive events in High energy Collisions at LHC energies.}

\author{ Sadhana~Dash}
\email{sadhana@phy.iitb.ac.in}
\affiliation{Indian Institute of Technology Bombay, Mumbai 400076, India}

\author{Nirbhay~Behera}
\email{nirbhay.iitb@gmail.com}
\affiliation{ Central University of Tamil Nadu, India}

\author{ Basanta~Nandi}
\email{basanta@phy.iitb.ac.in}
\affiliation{Indian Institute of Technology Bombay, Mumbai 400076, India}

\begin{abstract}
The charged particle multiplicity distribution in high energy hadronic and nuclear collisions receive contribution from both diffractive and non-diffractive processes. It is experimentally 
challenging to segregate diffractive events from non-diffractive events. The present work aims to separate and extract the charged particle multiplicity distribution of diffractive and non-diffractive events in hadronic collisions at LHC energies. A data driven model using the topic modelling statistical tool, DEMIX, has been used to demonstrate the proof of concept for p$-$p collisions at  $\sqrt s$ =  7 TeV generated by Pythia 8 event generator. The study suggests that DEMIX technique can be used to extract the underlying base distributions and fractions for experimental observables pertaining to diffractive and non-diffractive events at LHC energies and can therefore be used as a step forward for an experimental determination of  precise inelastic cross-sections in  p$-$p collisions.
\end{abstract}

\maketitle

\section{Introduction}

The estimation of total inelastic  proton$-$proton (p$-$p) cross-section is an important  observable to characterize the global properties of interactions. The inelastic cross-section receives a significant contribution ($\sim$ 26 - 29\%) from diffractive processes{alice1,atlas1,cms1}. It is imperative to estimate the diffractive contribution to the total inelastic cross-section as the diffractive cross-sections cannot be calculated in the pQCD framework and they depend on predictions made by models based on Regge theory \cite{regge}.
 The precise estimation of p$-$p inelastic cross-section and improving the theoretical description of diffractive processes is quintessential as it serves as an important input for modelling hadronic interactions in various  theoretical models and is  required for the calculation of the number of binary collisions in heavy ion physics.  

\begin{figure}
\includegraphics[width=1.0\linewidth]{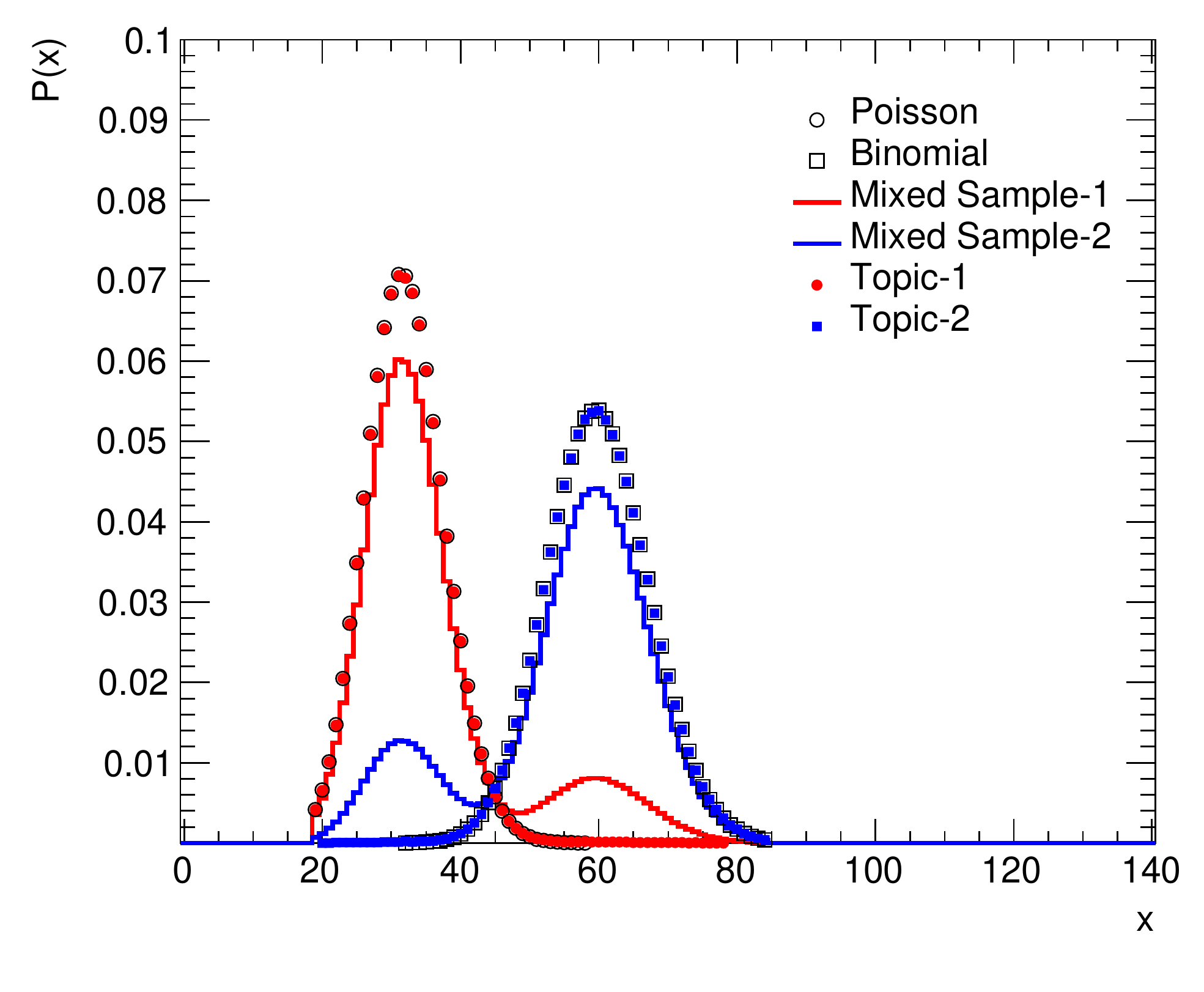}
\caption{ The  Poisson ( $\mu = 30$) and the  Binomial distribution function (60,55 ) used to generate the mixed samples by mixing them in different fractions (sample-1, 85:15 ; sample-2 , 18:82) in the toy model. The base distributions corresponding to Poisson and Binomial distributions are compared to the extracted topics. }
\label{fig1}
\end{figure}

The diffractive processes are characterized as dissociative processes with absence of net color flow between the initial state protons while non-diffractive ones have color flow between them. 
Generally, the diffractive events are characterized by exchange of pomerons which are color-singlet particles carrying the vaccuum quantum numbers. 
The current work is based on the DEMIX statistical tool  which has been successfully used recently to  separate quark and gluon jets  in p$-$p collisions \cite{jetprl, demix1}. Additionally the investigation has been extended to study differences in quark and gluon jet modifications in heavy ion collisions\cite{demix2}. The DEMIX method is commonly used to isolate mixture of  two different probability distributions to their base distributions \cite{topic1,topic2,topic3}. In this work, the method has been explained in terms of distributions functions related to diffractive and non-diffractive events.  Generally the method works for any two distinct classes of events.
Let  $p_{1}(x$) and $ p_{2}(x)$ be the probability distributions of charged particle multiplicity  in sample $S_{1}$ and $S_{2}$ respectively. They can be expressed as distinct sum of the underlying probability distributions of diffractive and non-diffractive events, $p_{D}(x)$ and $p_{ND}(x)$, respectively. These underlying pure distributions are known as the base distributions. 
Therefore, $p_{i}(x)$ can be expressed as \\
\begin{equation}
p_{i} (x) = c_{i} p_{D}(x) + (1 - c_{i}) p_{ND}(x)\\
\end{equation} 
where $c_{i}$s ($0 < c_{i} <1$) are fractions and one can have infinite ways of obtaining $p_{i}(x)$ depending on the values of $c_{i}$s.
A further constraint required for the tool to work is that one of the samples, $S_{1}$ should be enriched in diffractive events while the other sample, $S_{2}$ in non-diffractive events.  This requirement  enables one to use  the maximum likelihood ratio, $L_{{S_{1}}/S_{2}}$  as a classifier to distinguish between $S_{1}$ and $S_{2}$. The likelihood ratio  can be written as :
\begin{equation}
L_{S_{1}/S_{2}} (x) \equiv   \frac{p_{1}(x)}{p_{2}(x)} =  \frac{ c_{1} L_{D/ND}(x)  + (1 - c_{1})} { c_{2} L_{D/ND}(x)  + (1 - c_{2}) } 
\end{equation} 
where $L_{D/ND}(x) = \frac{p_{D}(x)}{p_{ND}(x)}$ is the likelihood ratio of diffractive and non-diffractive events in charged particle multiplicity space. As $L_{S_{1}/S_{2}} (x)$ is a monotonic function of  
$L_{D/ND}(x)$, one can find the diffractive enriched regions in $x$ where  $L_{D/ND}(x)$ is maximum. If one can find regions in $x$ observable  where the diffractive  and non-diffractive  distributions are pure, then  $L_{ND/D}(x_{D}) = 0$ or $L_{D/ND}(x_{ND}) = 0$.  One can then define the reducibility factor, $\kappa(D,ND)$ as \\
\begin{equation}
\kappa(D,ND) =  \min_{x}  \frac{p_{D}(x)}{p_{ND}(x)}  
\end{equation}  
If mutually irreducible in $x$,  $\kappa (p_{D} , p_{ND})$ = $\kappa (p_{D} , p_{ND})$=0.
Thus, $p_{1}(x$ and $ p_{2}(x)$ can be expressed in form of mutually irreducible distributions, $t_{D}(x)$ and $t_{ND}(x)$ for which $\kappa (t_{D} , t_{ND})$ = 0. These distributions  are referred as topics.  The $\kappa$s are related to the fractions $c_{i}$s. Therefore, the operational definition of diffractive and non-diffractive events can be considered as the topic distributions and are expressed as \\
\begin{eqnarray*}
t_{D}(x) =   \frac{p_{1}(x)  - \kappa(D, ND,) p_{2} (x)}{ 1 - \kappa{}D,ND} \\
t_{ND}(x) =   \frac{p_{2}(x)  - \kappa(ND, D,) p_{1} (x)}{ 1 - \kappa{}ND,D} \\
\end{eqnarray*} 
The details of the method are given in \cite{demix1}.
The method was tested through a  toy model  where  two different  probability distributions are generated  assuming  Poisson ( mean = 30) and Binomial ( mean= 60, variance = 55) distribution, respectively. The  two mixed samples were constructed  by mixing the distributions  such as the fraction of Poisson function in one distribution (sample-1) is 85\% and in the second sample (sample-2 ) is 18\%. The DEMIX procedure was then applied to obtain the topic distribution and was compared to the generated base distribution. This is shown in Figure \ref{fig1} and one can observe that the base distributions are in excellent agreement  with the extracted topics. The extracted fraction of Poisson function  in sample-1 after applying DEMIX  is 0.85 while it is 0.18 in sample-2. 

\begin{figure}
\includegraphics[width=1.0\linewidth]{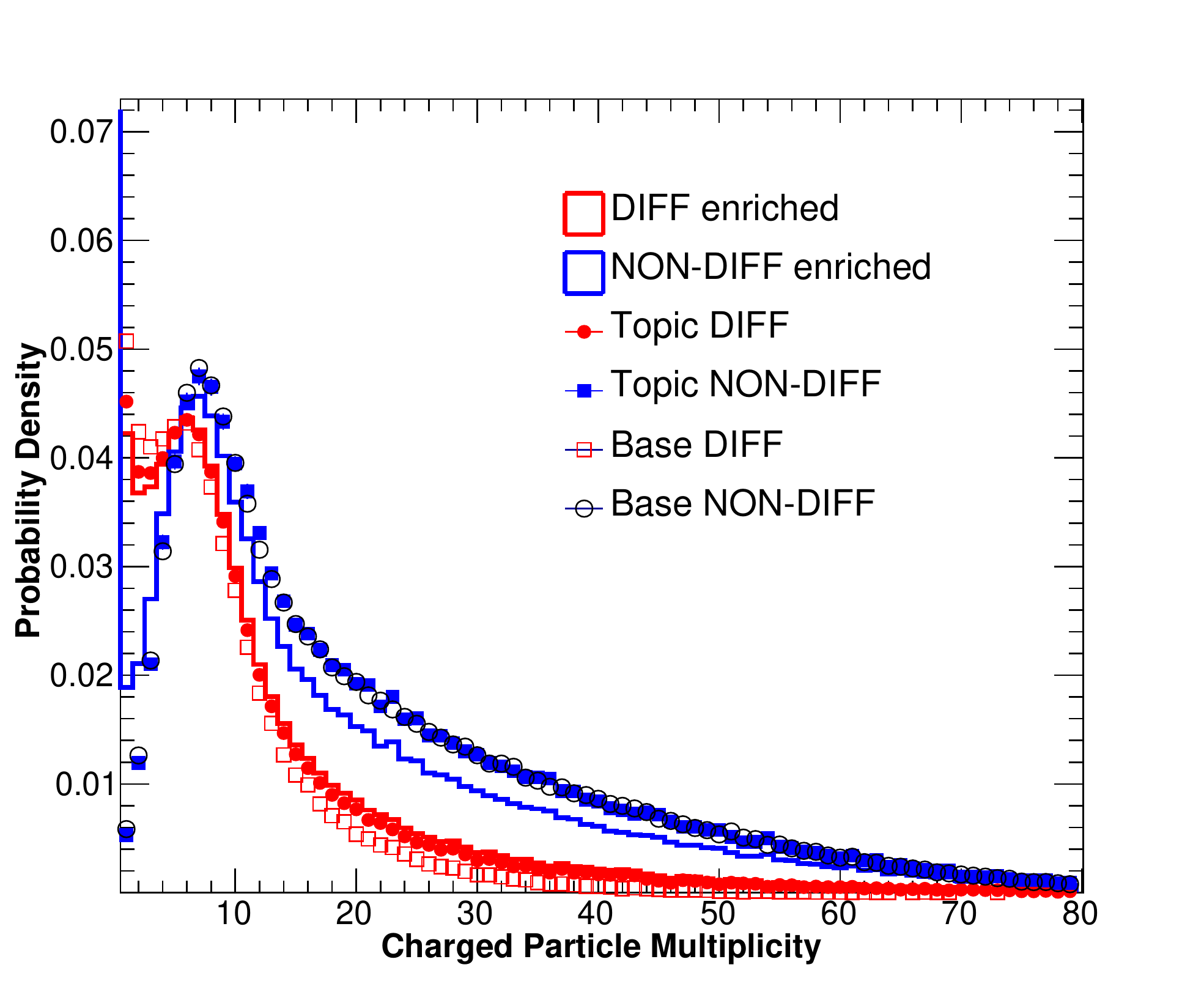}
\caption{ The distribution of charged particle multiplicity in samples enriched with diffractive and non-diffractive events for p$-$p collisions at $\sqrt{s}$ = 7 TeV generated by Pythia 8.2 model. 
The base distributions corresponding to pure diffractive and non-diffractive events are also shown. The extracted topics corresponding to diffractive and non-diffractive events for charged particle multiplicity extracted by DEMIX statistical method.}
\label{fig2}
\end{figure}

It is quite challenging to differentiate between the diffractive and non-diffractive events at the event level in high energy hadronic (p$-$p) collisions. This study is basically a proof-of -concept  which demonstrates that DEMIX method can be used to separate the diffractive and non-diffractive components of charged particle multiplicity distribution. The method present in this work is similar to that employed in \cite{demix2} to separate quark and gluon jets. 
The diffractive and non-diffractive events were generated using the Pythia 8 event generator for p$-$p collisions at  $\sqrt{s}$ =  7 TeV. Pythia 8 event generator has been extensively used to study both diffractive and  non-diffractive physics at LHC energies \cite{pythia8}. The diffractive events were comprised of single diffractive, double diffractive and central diffractive events \cite{sas1,sas2}. 
The two mixed samples of charged particle multiplicity distribution, one enriched in diffractive events and other in non-diffractive events were obtained by mixing different fractions of the generated non-diffractive and diffractive events. The diffractive enriched sample (DIFF) has  80 \% diffractive events and 20\% non-diffractive events while non-diffractive enriched events (NON-DIFF) has 80\% non-diffractive events. The charged particles were required to have $|\eta| < 2.5$.
Figure  \ref{fig2} shows the diffractive and non-diffractive topic distribution extracted from the DEMIX method. The extracted topics are compared to the base distributions of charged particle multiplicity for diffractive and non-diffractive events. One can observe that the extracted topics are in good agreement with the base distributions obtained by Pythia 8.2. The obtained $c_{2}$ = 0.22 which is 
close to the  input fraction of 0.20 used for generating the mixed sample. 

\begin{figure}
\includegraphics[width=1.0\linewidth]{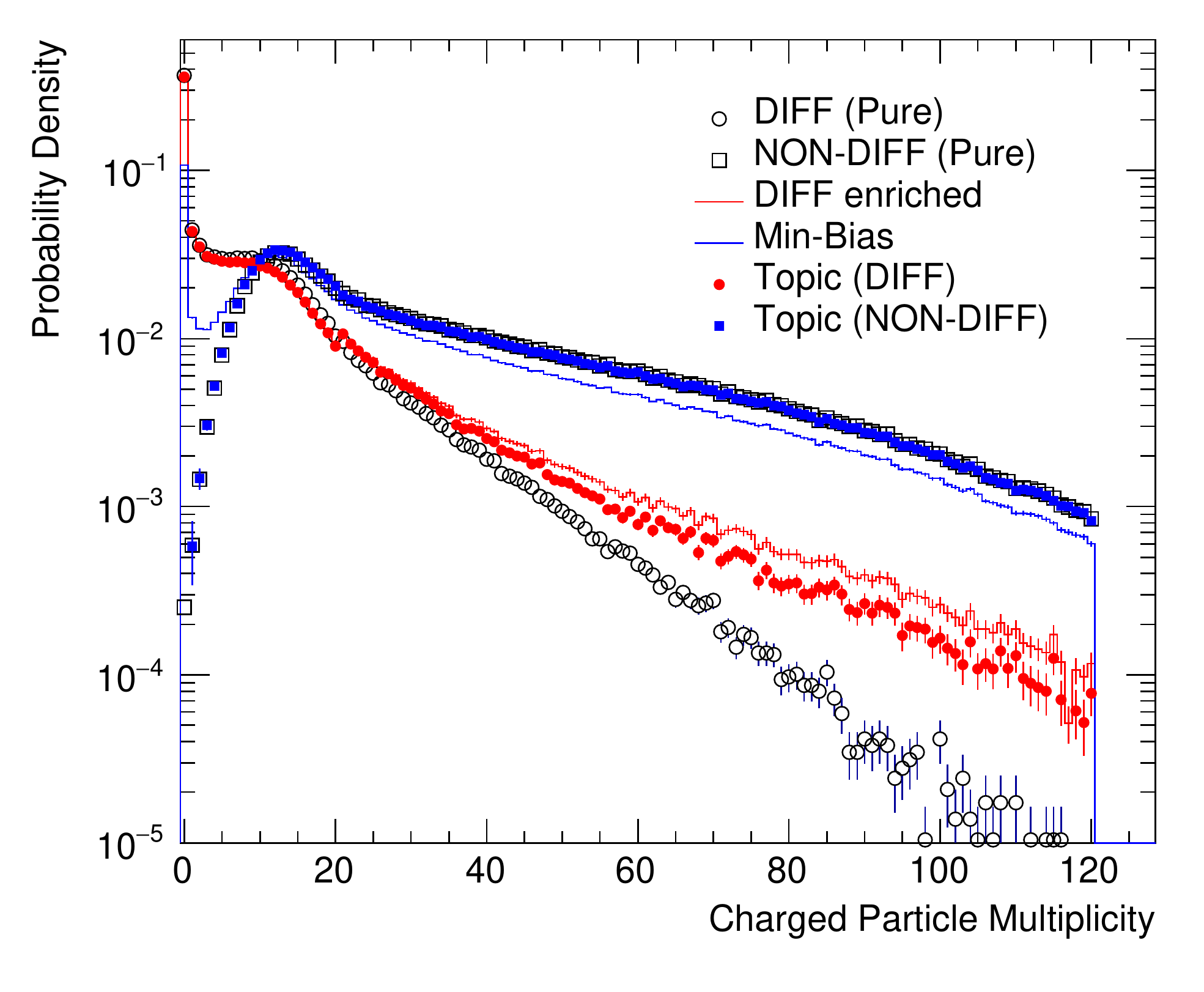}
\caption{ The distribution of charged particle multiplicity in samples enriched with diffractive and non-diffractive events (minimum bias sample) for p$-$p collisions at $\sqrt{s}$ = 7 TeV generated by Pythia 8 model. The base distributions corresponding to pure diffractive and non-diffractive events are also shown. The extracted topics corresponding to diffractive and non-diffractive events for charged particle multiplicity extracted by DEMIX statistical method.}
\label{fig3}
\end{figure}

However, in realistic experimental scenario, one does not have prior information about the enriched samples at event level and one has to depend on various kinematic cuts dictated by the diffractive physics like presence of large rapidity gap, diffractive mass etc at event level to segregate the diffractive  events. 
The recent  estimation of inelastic cross-section in p$-$p collisions at $\sqrt{s}$ = 7 TeV by  ATLAS experiment  used the DL model \cite{dlmodel}  with Pythia 8 fragmentation for calculating diffractive cross-sections and uncertainties\cite{atlas1}. Therefore, the present study can provide a model independent estimation of the cross-sections in real data. The study has been extended to minimum bias sample obtained with Pythia 8 to differentiate between diffractive and non-diffractive charged particle distributions.  The default minimum bias Pythia 8 events has  $\sim$ 29\% diffractive events and  $\sim$ 71\% non-diffractive events and is quite similar to the real experimental data. The sample-1 was taken to be the DIFF enriched  while the minimum bias sample was used as sample-2  for this study. The charged particles were required to have $|\eta| < 2.5$ and $p_{T} > 0.05$ . Figure \ref{fig3} shows the charged particle multiplicity distributions of the enriched samples, the topics extracted and their comparison to the base distributions in the accepted region. It can be observed that the distribution of topics extracted from the two distributions are in good agreement with the base distributions in the selected kinematic range.  The  fraction of diffractive events in min-bias was obtained to be 29.1\% which is in agreement with  the  default  fraction of $0.29$ present in the minimum bias Pythia sample.  This shows that one can use this technique in high energy hadronic collisions to obtain diffractive cross-section for precise measurements of inelastic cross-section.

\section{Summary}. 
The charged particle multiplicity distributions in high energy hadronic and nuclear collisions receive contribution from both diffractive and non-diffractive processes and it is imperative to distinguish them to obtain precise inelastic cross-sections. A data driven model using the statistical tool DEMIX was used to extract the fraction of  diffractive component in the charged particle multiplicity distribution of p$-$p collisions at LHC energies. The application of tool suggests that DEMIX technique can be used to extract the underlying base distributions and fractions for experimental observables pertaining to diffractive and non-diffractive physics at LHC energies and therefore can be used for a model independent experimental determination of inelastic cross sections to a better precision.

\section{Acknowledgements}
The authors would like to thank the Department of Science and Technology (DST), India for supporting the present work.


\begin{thebibliography}{50}

\medskip

\bibitem{regge} S. Donnachie, Hans Gunter Dosch, O. Nachtmann, and P. Landshoff, {\bf Pomeron Physics and QCD },Cambridge
University Press, Cambridge, England, (2002).
\bibitem{alice1} B. ~Abelev {\it et. al}, ALICE Collaboration, Eur. Phys. J. {\bf C 73}, 2456 (2013).
\bibitem{atlas1} ATLAS Collaboration, Nat. Commun. {\bf 2}, 463 (2011).
\bibitem{cms1} V. Khachatrayan {\it et. al}, CMS collaboration,  Phys. Rev. {\bf D 92}, 012003 (2015).
\bibitem{jetprl} Eric M. Metodiev and Jesse Thaler, Phys. Rev. Lett. {\bf 120}, 241602 (2018).
\bibitem{demix1} P. T. ~Komiske, Eric M. Metodiev and Jesse Thaler, JHEP {\bf 2018}, 59 (2018).
\bibitem{demix2} J. ~Brewer, J. ~Thaler and Andrew Turner,  Phys. Rev.{\bf C 103}, L021901 (2021).
\bibitem{topic1} David M. Blei, Coomunications of the ACM, {\bf 55},4,77-84 (2012).
\bibitem{topic2} Julian Katz-Samuels , Gilles Blanchard and Clayton Scott, Journal of Machine Learning research,  {\bf 20(41)},1-57, (2019).
\bibitem{topic3} J Ahn {\it et. al}, Bioinformatics {\bf 29}, 15, 1865-1871 (2013).
\bibitem{sas1} Gerhard. ~A. Schuler and Torbjorn Sjostrand , Phys. Rev. {\bf D 49}, 2257 (1994).
\bibitem{sas2} Gerhard. ~A. Schuler and Torbjorn Sjostrand , Z. Phys. {\bf C 73}, 677 (1997).
\bibitem{pythia8} Torbj{\"o}rn Sj{\"o}strand, Stefan Ask, Jesper~R Christiansen, Richard Corke, Nishita Desai, Philip Ilten, Stephen Mrenna, Stefan Prestel, Christine~O Rasmussen, and Peter~Z Skands, Comput. phys. commun. {\bf 191}, 159--177 (2015).
\bibitem{dlmodel} Donnachie, A. and Landshoff, P. V, Nucl. Phys. {\bf B 244}, 322–340 (1984).






\end{thebibliography}
\end{document}